\documentclass[a4paper,twoside,10pt]{letter}
\usepackage{graphicx,saj,multicol,subeqnarray,amsmath,url}

\def\papertype{Original Scientific Paper}

\def\p0{\phantom{0}}
\def\it{\sl}

\def\udc{...}
\begin{document}
\baselineskip=3.1truemm
\columnsep=.5truecm
\newenvironment{lefteqnarray}{\arraycolsep=0pt\begin{eqnarray}}
{\end{eqnarray}\protect\aftergroup\ignorespaces}
\newenvironment{lefteqnarray*}{\arraycolsep=0pt\begin{eqnarray*}}
{\end{eqnarray*}\protect\aftergroup\ignorespaces}
\newenvironment{leftsubeqnarray}{\arraycolsep=0pt\begin{subeqnarray}}
{\end{subeqnarray}\protect\aftergroup\ignorespaces}
%


\markboth{\eightrm MASS EXTINCTION AND THE STRUCTURE OF THE MILKY WAY} {\eightrm 
M. D.~FILIPOVI{\' C}, J.~HORNER, E. J. CRAWFORD, N. F. H. TOTHILL}

{\ }

\publ

\type

{\ }


\title{MASS EXTINCTION AND THE STRUCTURE OF THE MILKY WAY}


\authors{M. D.~Filipovi\'c$^1$, J.~Horner$^{2,3}$, E. J. Crawford$^1$, N. F. H. Tothill$^1$ }

\vskip 3mm


\address{$^1$University of Western Sydney, Locked Bag 1797, Penrith South DC, NSW 1797, Australia}
\Email{m.filipovic}{uws.edu.au n.tothill@uws.edu.au e.crawford@uws.edu.au}
\address{$^2$School of Physics, University of New South Wales, Sydney 2052, Australia} 
\address{$^3$Australian Centre for Astrobiology, University of New South Wales, Sydney 2052, Australia}
\Email{j.a.horner}{unsw.edu.au}


\dates{August 2013}{September 2013}


\summary{We use the most up to date Milky Way model and solar orbit data in order to test the hypothesis that the Sun's galactic spiral arm crossings cause mass extinction events on Earth. To do this, we created a new model of the Milky Way's spiral arms by combining a large quantity of data from several surveys. We then combined this model with a recently derived solution for the solar orbit to determine the timing of the Sun's historical passages through the Galaxy's spiral arms. Our new model was designed with a symmetrical appearance, with the major alteration being the addition of a spur at the far side of the Galaxy. A correlation was found between the times at which the Sun crosses the spiral arms and six known mass extinction events. Furthermore, we identify five additional historical mass extinction events that might be explained by the motion of the Sun around our Galaxy. These five additional significant drops in marine genera that we find include significant reductions in diversity at 415, 322, 300, 145 and 33~Myr ago. Our simulations indicate that the Sun has spent $\sim$60\% of its time passing through our Galaxy's various spiral arms. Also, we briefly discuss and combine previous work on the Galactic Habitable Zone with the new Milky Way model.}



\begin{multicols}{2}


\section{1. INTRODUCTION}

Mass extinctions have the effect of wiping the biological slate clean, freeing up ecological niches and thus producing explosions in biodiversity (e.g. McElwain and Punyasena 2007; Alroy 2008). In the past, several explanations have been proposed to resolve ancient mass extinctions, including vast outpourings of flood basalt (such as the Deccan and Siberian Traps; e.g. Wignall~2001), periods of global glaciation (Mayhew et al.~2008) and the impact of large asteroids and comets upon the Earth (e.g. Alvarez et al. 1980; Bottke et al.~2007). Of these, extreme geological and climate phenomena such as flood basalt outpouring and ``snowball Earth'' glaciations appear to be very rare and randomly-occurring events in the Earth's history. Overholt et al. (2009) investigate Earth's climate as a function of location in the Galaxy, however, no obvious correlation could be drawn. On the other hand, it is well established that the Earth has been continually pummelled by asteroidal and cometary impactors throughout its history, a process that will continue well into the future. Given the damage that would be caused by the impact of a cometary or asteroidal body several kilometres in diameter (an event expected to reoccur on timescales of millions or tens of millions of years), it seems like that the majority of mass extinctions could be caused by such impact events. Thus, impact theories are strong contenders to explain mass extinctions.

In addition to the extinction risk due to impactors, there is also the possibility that nearby supernovae could also cause mass extinctions. In this case, one would expect the nearby supernova flux to be higher whilst the Sun is traversing the Galaxy's spiral arms than when it is between them (Svensmark 2012). However, given the low frequency of supernovae, the likelihood of one occurring sufficiently close to the Earth to trigger a mass extinction is thought to be relatively low, even during spiral arm crossings (Beech~2011). As such, here, we assume that collisions with comets and asteroids are the dominant cause of exogenous mass extinctions (i.e. those extinctions whose cause is external to the Earth).

The hypothesis of mass extinction driven by cometary or asteroidal impact is part of the burgeoning modern interdisciplinary study of astrobiology -- a field in which researchers from the breadth of all the natural sciences come together to try to understand the origin, diversity and history of life on Earth and the prospects for life beyond our Solar system (e.g. Horner and Jones 2010). When considering life on Earth, biologists and geologists have long reported evidence for mass extinctions throughout the history of our planet (e.g. Horner et al.~2009), but have found it difficult to find explanations for those extinctions. Astronomical studies not only inform biologists on the conditions that would have been experienced by the earliest life on the planet, and of the origins of the water considered so vital for life to develop and thrive (e.g. Horner et al.~2011), but can also be used to attempt to explain those mass extinctions for which a terrestrial cause remains elusive.

One of the most intriguing suggestions related to the mass extinctions on Earth is that those extinction events are not randomly distributed through time. Instead, a number of authors have suggested that there is a periodic signal within the mass extinction record, with both the historical major mass extinctions and a number of more minor extinction events following a periodic pattern. In recent years, a number of studies (Rohde and Muller 2005; Melott and Bambach 2011, 2013; Feng and Bailer-Jones 2013) have discussed and analysed a proposed $\sim$62~Myr period.

 \subsection{1.1. Impacts on the Earth}

The meteors which can be observed on any clear night represent the small, non-threatening end of a spectrum of regular impacts. The largest and most devastating impacts are the least frequent, whilst the smallest (meteors) are so frequent that millions occur across our planet every day. The Earth Impact Database\footnote{http://www.passc.net/EarthImpactDatabase/index.html} currently lists a total of 182 confirmed large impact structures across our planet's surface. These structures are the scars left behind as a result of collisions between the Earth and asteroidal or cometary objects, and represent just a tiny fraction of the true impact history of our planet. The majority of impacts occur in the Earth's oceans (which make up $\sim$70\% of the planet's surface area) and therefore, despite likely causing devastation at the time, rarely create scars which would survive to the current day to be analysed. Indeed, studies show that a layer of water can significantly reduce the ability of an impactor to leave an impact crater on the ocean floor (e.g. Baldwin et al. 2007; Milner et al. 2008). Since the average depth of the oceans is $\sim$6 km, it is clear that the great majority of impacts will fail to leave any recognisable scar on the ocean bottom. Furthermore, the ocean floor is recycled on timescales far shorter than the age of our planet, effectively erasing any evidence of ancient impact scars. Even for those impacts which occur on land, erosion and weathering remove the scars from all but the largest impacts and astronomically short timescales. A true idea of the ongoing impact regime experienced by the Earth is therefore best obtained by looking at our nearest neighbour, the Moon, or by examining the surface of Mars (where the effects of weathering and erosion are far less effective at removing the scars left behind by impacts of all scales). Both the Moon and Mars are far more heavily scarred than the Earth -- and both display evidence that impacts are certainly a current, rather than historical, concern. The repeated impacts that have been observed on the giant planet Jupiter over the last twenty years add further weight to this argument -- both the large Shoemaker-Levy~9 impacts in 1994 (Hammel et al.~1995); and the smaller impacts observed in the last few years (e.g. S\'anchez-Lavega et al. 2010).

Over the years, there have been many suggestions that the impact flux of such asteroids and comets upon the Earth has varied significantly as a function of time. The rate of the smallest impacts (i.e. meteors) appears to vary periodically throughout the course of the year, as the Earth encounters streams of debris left behind by passing asteroids and comets\footnote{A simple illustration of this variation can be found in the annual Meteor Shower Calendar hosted by the International Meteor Organisation -- http://www.imo.net/calendar/2013}. For the larger, more threatening impacts, too, there are suggestions of periodicity (e.g. Raup and Sepkoski 1986; Rampino and Stothers 1984; Rampino 1997; Chang and Moon 2005) -- although it is hard to uncover a clear result because of the small number statistics involved in the study of such events. The most widely-accepted hypothesis of time-variation in the record of massive impacts is that of the ``Late Heavy Bombardment''.

 \subsection{1.2. The Late Heavy Bombardment}

The most widely discussed example of temporal variation in the Earth's impact flux is the Late Heavy Bombardment. This hypothesis suggests that, early in the history of the Solar System, the Earth and Moon were subjected to so many massive impacts as to make the Earth entirely uninhabitable (e.g. Oberbeck and Fogleman 1989; Grieve and Pesonen 1992; Gogarten-Boekels et al. 1995; Wells et al. 2003), due to the repeated sterilisation of the planet. The proposed bombardment, thought to have continued until around 800 million years after the Earth's formation, is thought to have been linked to the creation of the ``seas'' on the Moon. 

Current ideas of the Late Heavy Bombardment (e.g. Gomes et al. 2005; Levison et al. 2008) suggest that it was a side-effect of the migration of the giant planets (Jupiter, Saturn, Uranus, Neptune) in the early Solar system. In their models, the initial architecture of the orbits of the giant planets was significantly more compact than that we observe today, with a large amount of material located just beyond the orbit of the outermost planet. As Jupiter and Saturn migrated, they eventually reached a regime where their orbits were strongly mutually interacting, which resulted in the chaotic evolution of the orbits of all four giant planets. In the models presented by those authors, this resulted in the outward scattering of Uranus and Neptune into the massive disk of planetesimals that lay beyond, dispersing that disk (and circularising the orbits of those planets), and in the process flinging vast amounts of cometary and asteroidal material towards the terrestrial planets. The result was a short but remarkably intense period of cataclysmic impacts on the terrestrial planets -- the Late Heavy Bombardment, following which the Solar System would have relaxed to its current relatively quiescent state. However, Norman (2009) strongly questioned the cataclysm hypothesis, pointing out a string of inconsistencies in establishing absolute ages of ancient impact basins and the sources for the impactors.

There are other theories that relate the rate of impacts on Earth to the dynamics of our Solar system, many of them focusing on the possible presence of a companion body to the Sun, which perturbs the orbits of comets and planetary debris enough to put them on a collision course with the Earth (e.g. Davis et al. 1984; Whitmire and Jackson 1984; Matese et al. 1995, 1999; Horner and Evans 2002; Matese and Whitmire 2011; Sumi et al. 2011). 

Whilst the idea of the Late Heavy Bombardment is still heavily debated (e.g. Haskin et al. 1998; Chapman et al.~2007), the hypothesis clearly demonstrates the importance of understanding astrobiological events and their consequent impact on the biological development of life -- hence the field of astrobiology. Theories of planetary migration (themselves inspired both by our study of extrasolar planetary systems and studies of the small bodies in our Solar system (e.g. Lykawka et al. 2009; Malhotra 1995; Nesvorn\'y et al. 2013; Minton and Malhotra 2011)) are used to explain how such a phenomenon could come about, and we come to understand that the planetary environment in which we arose is intimately connected with the detailed dynamical history of the Solar system.

 \subsection{1.3. Impact History and the Structure of Our Galaxy}

An interesting alternative explanation for the potential periodicity observed in the Earth's impact and extinction history is that the variation is the direct result of the periodic passage of our Solar system through the spiral arms of the Galaxy (see Fig.~1). Our Solar system lies at significant distance ($\sim$8~kpc) from the centre of a large spiral galaxy, the Milky Way. The Galaxy consists of a central bulge (far interior to the orbit of the Sun about its centre) surrounded by a number of spiral arms. The arms themselves contain large quantities of gas and dust, from which new generations of stars are continually being formed. The most massive stars, which are also the most short-lived, are heavily concentrated within the spiral arms, whilst the spaces between the arms are significantly less densely populated, being relatively free of gas, dust, and massive stars. Because the most massive stars have the shortest lives, supernovae (the cataclysmic explosions of the most massive stars) are also concentrated in spiral arms whose thickness may be up to $\sim$1-2~kpc (McClure-GrifÞths et al. 2004). 

The motion of the spiral arms around the centre of the Galaxy is somewhat slower than that of the stars that make up the galaxy, which means that, as the Sun orbits the centre of the Galaxy, it follows a path that takes it through the spiral arms every few tens of millions of years. In the spiral arm environment, the Solar System is exposed to a far more hazardous and busy regime than in the inter-arm regions (our current location). The Earth could be relatively close to a star when its life comes to an end in a supernova explosion -- which could certainly pose problems for life, although such supernovae are relatively rare, and the odds of the Earth being sufficiently close to one for life to be exterminated entirely are low, even within a spiral arm (Beech~2011). At the same time, close encounters between the Sun and neighbouring stars become more frequent, as do encounters between the Sun and giant gas clouds (see Fig.~2). Such encounters would not pose a direct hazard to life on Earth by changing the orbit of the Earth around the Sun, but could pose a hazard by disturbing the Oort Cloud (Porto de Mello et al.~2009), a vast cloud of comets (Oort~1950) which stretches to a distance of at least 100\,000~AU from the Sun. The Oort Cloud is thought to contain trillions of cometary nuclei, left over from the formation of the Solar system, which are only tenuously gravitationally bound to the Sun (the outer members of the cloud are around halfway to the nearest star). An encounter with a passing star or distant molecular cloud can be enough to deflect an Oort cloud comet, throwing it onto a new orbit that will bring it into the inner Solar system -- where it can pose a threat to the Earth. The closer the star approaches to the Sun, or the more massive it is (or both), the more comets it will scatter inwards, and therefore the more likely it will be that one of those in-falling comets will hit the Earth.

\end{multicols}

\centerline{\includegraphics[width=0.85\textwidth]{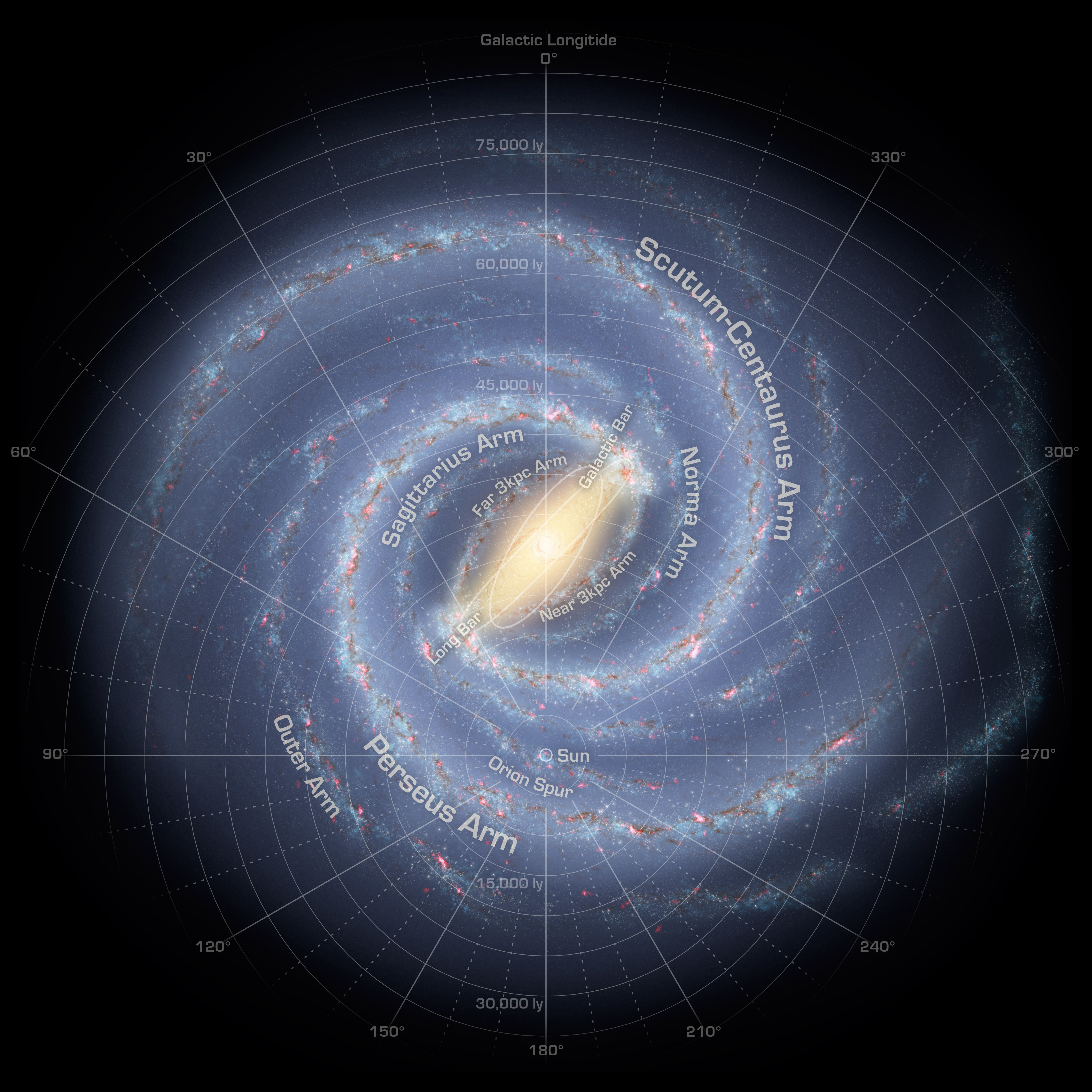}}
\figurecaption{1.}{The face-on view of the Milky Way (Churchwell et al. 2009). The location of the Sun is indicated, along with the names and locations of the spiral arms and spur. }

\begin{multicols}{2}

Such comet showers are not merely hypothetical -- the catalogue of observed long period comets (which come from the Oort cloud) contains a relatively weak, but still statistically significant, sign of a comet shower which is thought to have peaked a few million years ago. This shower (the Biermann shower) was first identified some thirty years ago, and illustrates how even relatively distant encounters between the Sun and passing stars can influence impacts on the Earth (Biermann et al. 1983).

Here, we test the idea that the Sun's orbit around the centre of the Milky Way has a significant influence on the impact regime experienced by the Earth. Using the latest results on the structure of our Galaxy, we construct a detailed and accurate timeline of the Sun's motion through spiral arms, and compare it to the latest knowledge of the history of biodiversity on Earth over time and the vastly improved dataset of global impacts that has become available over the last few years. Correlations between spiral arm crossings and mass extinctions suggest that the history of life on Earth is intimately connected with our place in the Universe.

\section{2. DATA AND MODELLING}

We make use of the major international impact databases to acquire the largest possible database of impact dates and sizes over the past billion years or so. We combine this with the latest understanding of the history of life on Earth, using the results of recent studies of the biodiversity of our oceans as a function of time to determine whether there is any correlation between that biodiversity and the flux of impacts on the Earth. Whilst it is true that any correlation between the two will be somewhat masked (by endogenous causes of mass extinctions, such as snowball Earth epochs, flood basalt outpourings, etc.), there are enough data to definitively determine whether the history of life on Earth has periodically been truncated by asteroid and cometary impactors. However, the further back we look, the more evidence is likely missing, lost to the ages past (i.e. as we go further back, there is less of a fossil record, so smaller mass extinctions could be missed/overlooked).

 \subsection{2.1. Models of the Sun's Trajectory through the Milky Way}

In recent years, our understanding of the structure of our Galaxy has improved dramatically. As a result of a number of highly detailed surveys, a new picture is emerging which reveals our Galaxy's structure in far more detail than has ever been shown before. With that model, it is possible to accurately estimate the timing of the Sun's last orbit through the Galaxy's spiral arms -- yielding timings that we can compare to the observed impact cratering record and extinction records to determine whether any correlation can be seen. 

The kinematic parameters of the Milky Way used in this work are based on previous estimates (Gies and Helsel 2005), with the angular velocity of the Sun set at 26.3~km~s$^{-1}$~kpc$^{-1}$ and the difference between the velocity of the Sun's motion and that of the spiral arms being 11.9~km~s$^{-1}$~kpc$^{-1}$ (Overholt et al. 2009), which means that the Sun moves at a significantly greater velocity than the spiral arms. According to these values, during the last 500~Myr, the Sun has almost completely circumnavigated the entire spiral arm pattern, crossing two major and two minor arms, and one or two interarm spurs. Svensmark (2006) estimated that the last two
spiral arm crossings happened approximately 31~Myr and 142~Myr ago. He also estimated that the spiral arm/interarm density ratio is in the
range 1.5--1.8.

\end{multicols}

\centerline{\includegraphics[width=0.5\textwidth]{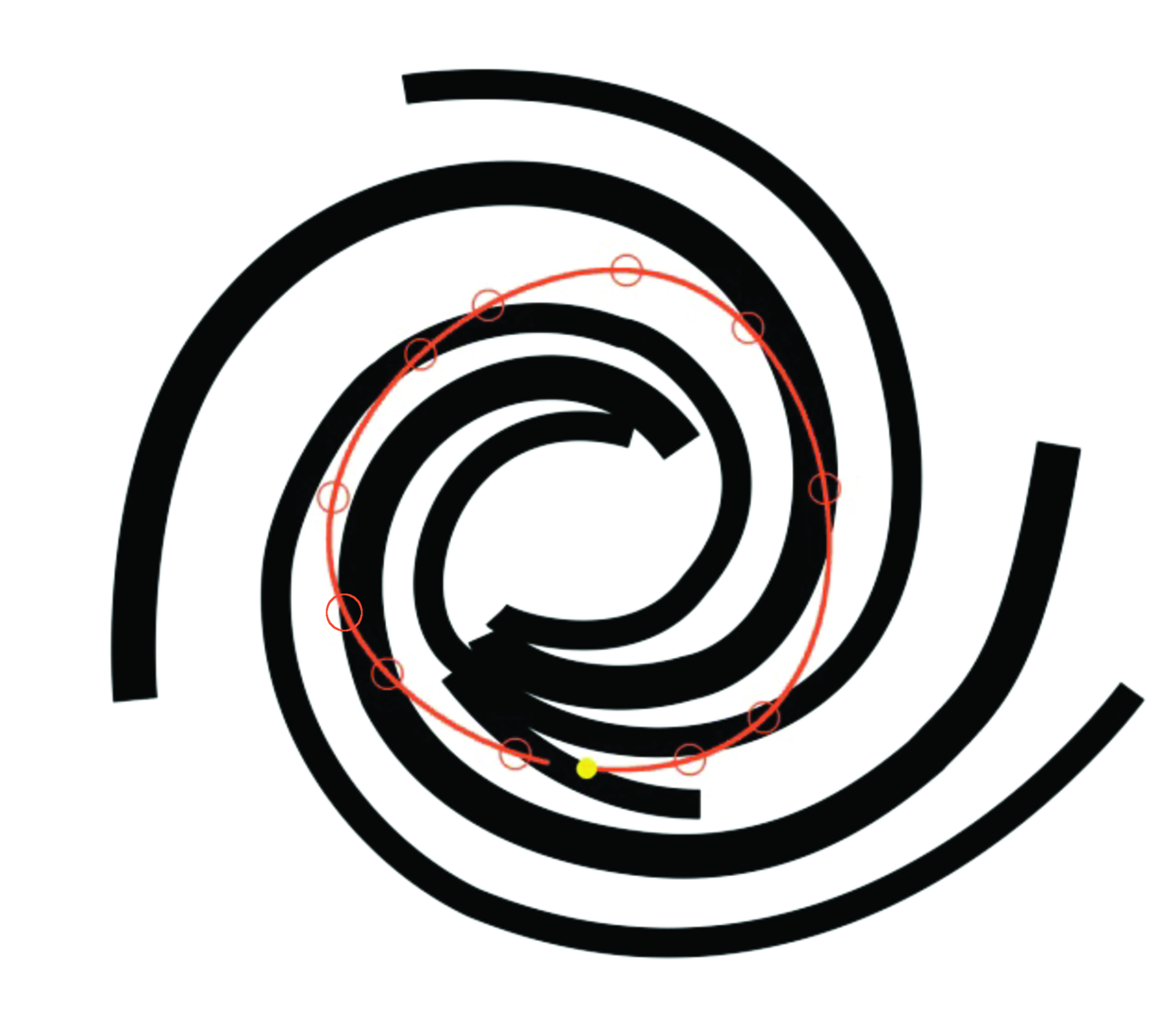}\includegraphics[width=0.5\textwidth]{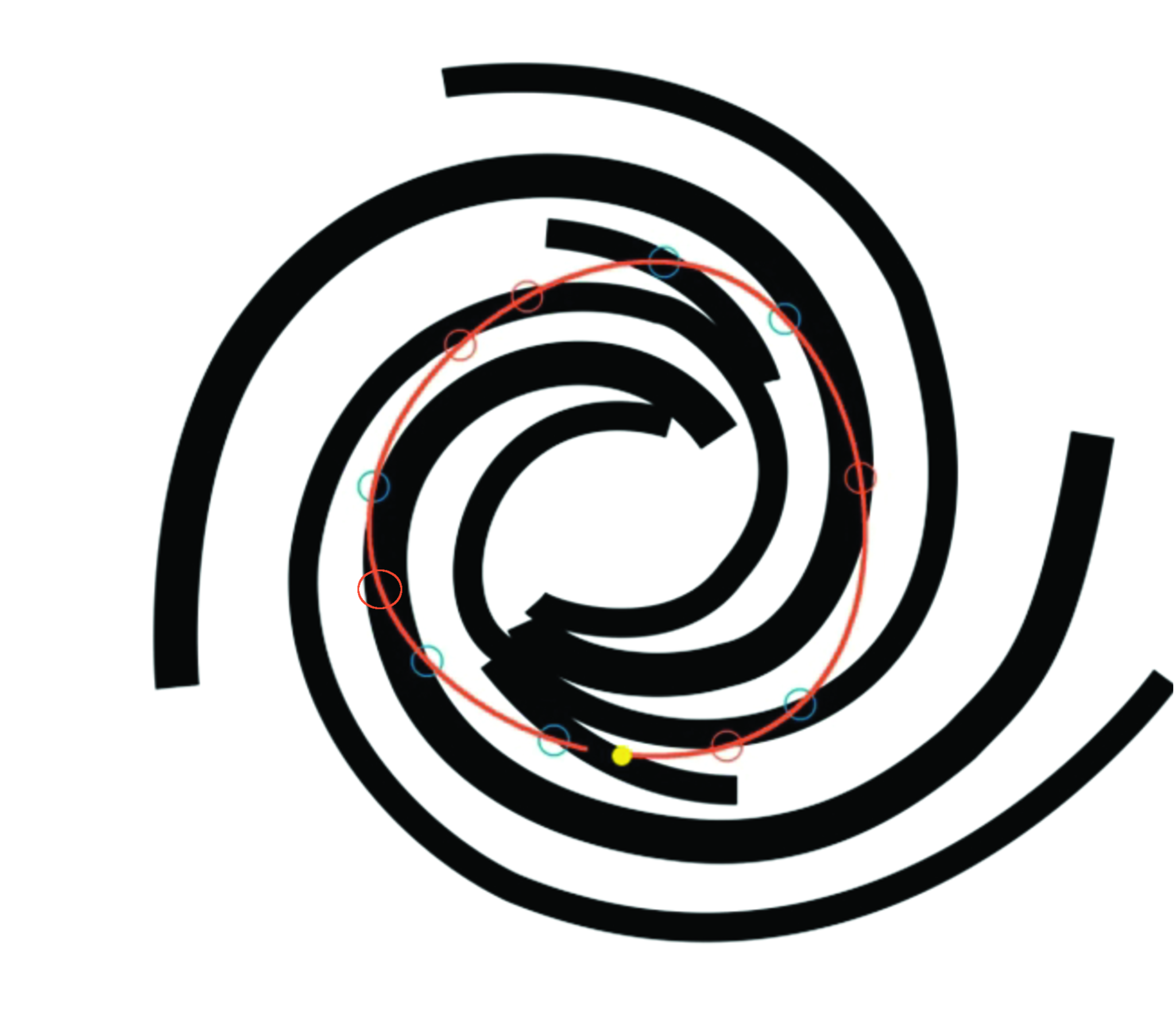}}
\figurecaption{2.}{The Milky Way model (left), based on Churchwell et al. (2009) and our new weighted symmetrical model (right) of the Galaxy with the Sun's orbital path over 500~Myr. The Sun's current position is indicated with a yellow dot. Eleven extinction events are shown along the path by circles. The six blue circles in the new Galaxy model represent the known mass extinctions as marked in Fig.~3, whilst the orange circles represent the five additional events we propose here. The thickness of two major arms is set to 1.5~kpc and minor arms (as well as spurs) to 1~kpc.}

\begin{multicols}{2}

Historically, it has been assumed that the Sun's crossing of the Galaxy's spiral arms is a relatively simple, periodic event. However, the new model of our Galaxy's structure (Figs.~1 and 2) reveals the truth to be significantly more complicated. The spiral arms are not evenly spaced, and a number of smaller sub-arms are dotted between them. The result is that encounters between the Sun and the spiral arms will be both more frequent and more randomly distributed in time. By using this new model of galactic structure, we are able to take account of this irregular behaviour for the very first time, enabling us to carry out the first fair study of the influence of the Galaxy's structure on the impact flux at Earth.

The model of the Sun's path through the Milky Way can be used as the basis for a model of the impact rate. Such a model would take account of the infall time for new comets (typically of order 500~kyr to 1~Myr), and feature gradual ramping up and slowing down of the impact flux at Earth, as the Solar system transitions between the low-flux state (whilst it lies between the spiral arms) to the high-flux state (whilst passing through the arms) and back to the low-flux state (as it returns to the space between the arms). During the time that the Solar system is located between the spiral arms of the Galaxy, encounters (whether with passing stars or giant molecular clouds) that might perturb the Oort cloud, causing a comet shower and an elevated impact flux will clearly be far less frequent than during the Sun's passage through the crowded spiral arms.

We also note that the model we are using and that of Gies and Helsel (2005) are similar in appearance, but that any differences would result in different gravitational potentials and thus slightly different trajectory shapes. We superimposed the Solar orbit from Gies and Helsel (2005) onto the Milky Way model based on Churchwell et al. (2009) (see Figs.~1 and 2). In Fig.~2, we mark the approximate locations of the known mass extinctions as a function of the Sun's orbit around the centre of the Galaxy, as described in the following section. 

Also, we include approximate observational limits on the location of those events that result from obscuration by material in the Galactic centre. This is mainly because of the observational constraints placed on our knowledge of the structure of the far side of our Galaxy.

 \subsection{2.2. Extinctions in Biodiversity}

The six largest mass extinction events of the last 500~Myr resulted in significant reductions to the biodiversity of the planet. These six extinction events are: Cretaceous-Paleogene (C-P) (also known as the Cretaceous-Tertiary or K-T, however the Tertiary period is not currently recognised by the International Commission on Stratigraphy) at 66~Myr ago; Triassic-Jurassic (T-J) at 200~Myr ago; Permian-Triassic (P-T) at 251~Myr ago; Late Devonian (LD) at 375~Myr ago; Late Ordovician (LO) at 445~Myr ago; Late Cambrian (LC) at 488~Myr ago. In Fig.~3 we display (blue vertical lines) the position of these mass extinction events along the geological timeline and the global marine genera number distribution over the last 500~Myr (Rohde and Muller 2005).

There is extensive coverage of these mass extinction events in the literature, therefore, we emphasise their position in time rather than their effect on ancient biodiversity. Our study is based on Sepkoski's definition\footnote{Sepkoski's online database can be accessed at http://strata.geology.wisc.edu/jack/} of a mass extinction event (Sepkoski~2002), which is a sharp decrease in marine genera along the diversity curve.

In addition to these mass extinctions, there are five other significant drops in marine genera that we find indicated in the data (they are also known in the literature as ``lesser extinctions''). These somewhat smaller reductions in marine genera diversity occur 415, 322, 300, 145 and 33~Myr ago (Fig.~3; orange vertical lines). Not all mass extinctions would be the result of spiral arm crossings. While other events could also contribute (the background impact flux due to the planetary system, and extinctions due to climate and geological reasons), we suggest that the probable cause for these reductions in marine genera is also connected to spiral arm crossings.

Although extinction events have consistently, and some would argue periodically, eroded away life on this planet, they have also likely driven the evolutionary radiation of the species. With the widespread loss of species, surviving species may fill new niches and physically adapt accordingly to new habitats. Without the mechanisms that brought about mass extinction events, the complexity of life as we know it today may not exist. If these mechanisms are of extraterrestrial origin, they may prove to be an essential ingredient, rather than a barrier, to complex life in the Galaxy.

As well as the addition of new extinctions we identify in the data, we also see a pattern in the marine genera data over the last 500~Myr. There is rapid growth beginning at 500~Myr, followed by a general decline in genera, then a prolonged growth leading to the present time. Because marine genera data, and for that matter data on land flora and fauna, are based on fossil evidence, these trends in the data may be due to the efficiency of fossilisation and possibly other unknown factors rather than actual genera populations over time (Bailer-Jones 2009). Bailer-Jones (2009) argues that there is no significant evidence for intrinsic periodicities in biodiversity, impact cratering or climate on timescales of tens to hundreds of Myr; therefore it seems likely that more than one mechanism has contributed to biodiversity variations over the past 500~Myr.

\end{multicols}

\centerline{\includegraphics[width=1\textwidth]{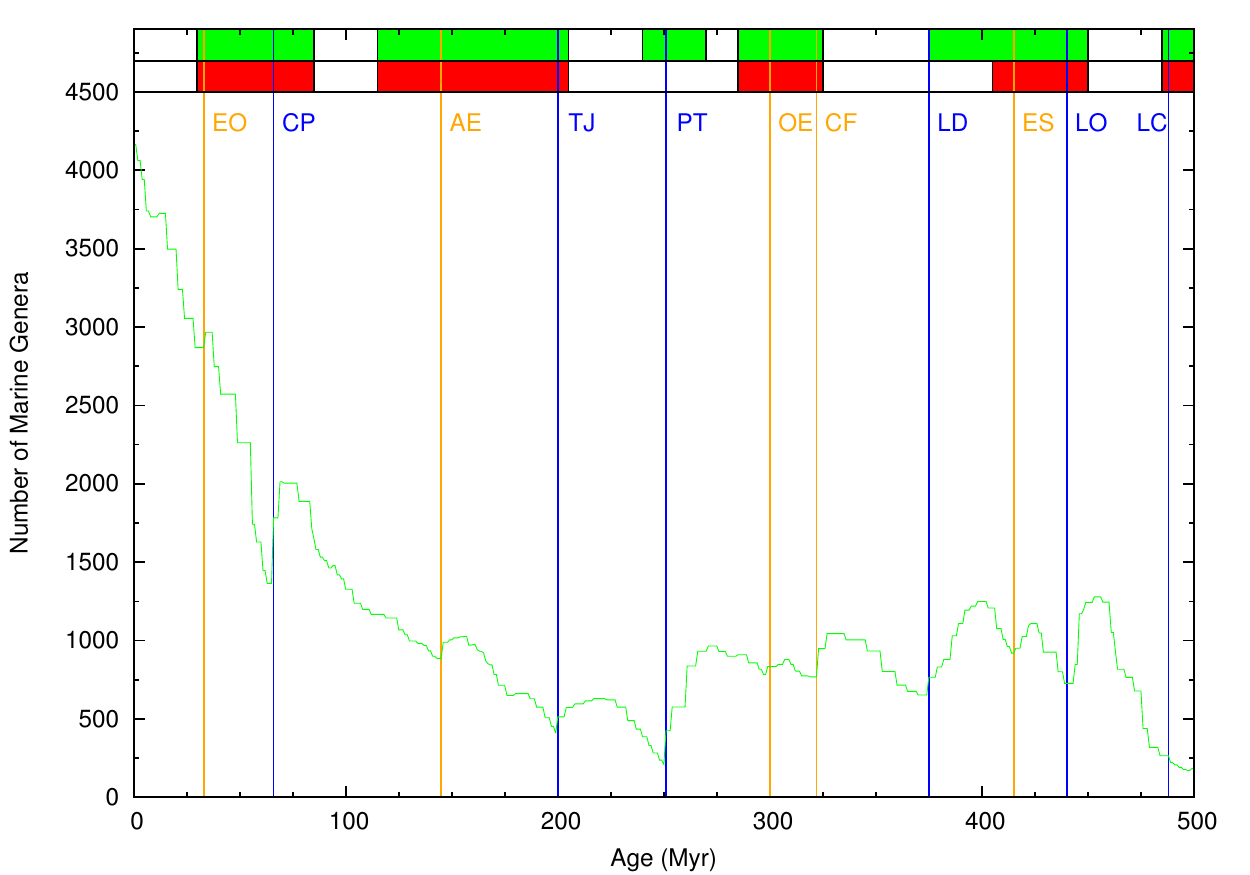}}
\figurecaption{3.}{The number of marine genera over geological time/age (in Myr). The major mass extinction events are indicated by blue lines; Cretaceous-Paleogene (CP; 66~Myr), Triassic-Jurassic (TJ; 200~Myr), Permian-Triassic (PT; 251~Myr), Late Devonian (LD; 375~Myr), Late Ordovician (LO; 445~Myr) and Late Cambrian (LC; 488~Myr). Five proposed new extinction events are indicated by orange lines; Eocene-Oligocene (EO; 33.9~Myr), Aptian Extinction (AE; 145~Myr), Olson's Extinction (OE; 300~Myr), Carboniferous Rainforest Collapse (CF; 322~Myr) and End-Silurian (ES; 415~Myr). The vertical lines correspond with the orange/blue circles in Fig.~2. The green shaded area on the top represent Sun's passage through the Milky way in the new model while in red we indicate passage in the older model. Figure adapted from Rohde and Muller (2005).}

\begin{multicols}{2}

\section{3. NEW MILKY WAY MODEL}

With a differential speed of 11.9~km~s$^{-1}$~kpc$^{-1}$, the Sun orbits the centre of the Milky Way galaxy, passing through four spiral arms and a spur, and almost completing one full circuit of the entire spiral arm pattern (last 500~Myr; see Fig.~2 (left)). The thickness of the major arms is set to 1.5~kpc and minor arms, as well as spurs, to 1~kpc.

We have identified eleven significant spikes in the extinction intensity data that we marked on the Sun's orbital path. If we were to contend that all eleven extinctions discussed herein were the result of our passage through the Galaxy's spiral arms, then it is clearly possible to modify our Galaxy model to account for those extinctions. Whilst this is a purely hypothetical exercise, it is not necessarily unreasonable. As can be seen in the right-hand panel of Fig.~2, only a small modification to the Milky Way model put forth by Churchwell et al. (2009) is required in order to ensure that all eleven mass extinctions occur within one or other of the spiral arms. This region of the Galaxy is far less known (observationally) and that significant level of symmetry does exist in other galaxies. 

In the original Milky Way model (Fig.~2 (left)), we observe that nine of the extinctions studied happened during the Sun's passage throughout the spiral arms. However, in our modified symmetrical model all eleven events are positioned in the spiral arms. Four (TJ, PT, OE and CF event) of these extinctions lie on a part of the Sun's orbit where our view of the Milky Way's structure is obscured by the Galactic centre. Therefore, this gives some freedom for speculation on a spur, similar to the Orion spur, existing on the other side of the Galaxy. If we were to assume that the extinctions are solely caused by impacts induced by our passage through spiral arms, then this would clearly allow us to constrain their locations independent of astronomical observations -- an interesting test for the current model of galactic structure. 

However, as we noted earlier, there are a number of endogenous factors that could also cause mass extinctions, and so it seems plausible to assume that those extinctions that fall outside of periods when the Sun is crossing a spiral arm could have an endogenic rather than an exogenic origin. Equally, as we also noted earlier on, being outside of a spiral arm does not preclude close encounters between the Sun and other stars, and so we would expect at least some impact induced extinctions to occur whilst the Sun is between crossings, albeit at a much reduced rate.

Summed over the entirety of its orbital path over the last 500~Myr, based on the new weighted symmetrical model, we find that the Sun has spent $\sim$60\% of its time in the spiral arms (see the green shaded areas at the top of Fig.~3). By comparison, for the case of the old non-symmetrical model, the Sun would have spent $\sim50\%$ of the same period in the spiral arms (red shaded areas in Fig.~3).

Given the incompleteness of the available data any claims on mass extinctions' temporal distribution are naturally highly speculative. 
We therefore employ a simple statistical prediction (null hypothesis) of exactly how many of the extinctions could occur randomly during spiral arm crossings, and estimate the likelihood of that occurring by chance. 

The null hypothesis is that no mass extinction has been caused by spiral arm crossings, and therefore all eleven mass extinctions occur within the spiral arms purely by chance. The probability of this is 0.36\% ($0.6^{11}$). The hypothesis that six out of eleven randomly occurring events fall inside spiral arms by pure chance has fairly high probability -- 20-25\% --  which is not negligible. However, the probability that all eleven random events fall within the spiral arms is low (0.36\%). 

A simple numerical test also produces similar results. We generated a series of data sets in which eleven extinction events were randomly distributed through the last 500~Myr, and then counted the frequency with which all eleven occurred during spiral arm crossings. Our results again support the hypothesis that it is highly unlikely for all of the studied mass extinctions to coincidentally fall during spiral arm crossings, suggesting that their timings are not simply a matter of chance.

Our results suggest that the proposed 62~Myr periodicity (Rohde and Muller 2005) in mass extinctions could be directly related to the Sun's passage through the spiral arms of the Milky Way. However, we stress that significant controversy about 62~Myr periodicity still exists among the scientific community (for detailed analysis see Feng and Bailer-Jones (2013) as well as Melott and Bambach (2013).

\section{4. CONCLUSIONS AND FUTURE WORK}

We created a new model of the Sun's orbit around the centre of the Milky Way, in order to accommodate the influence of spiral arm crossings on the cometary flux through the inner Solar system. Our model reveals the periods when the Earth has suffered the highest risk of cometary impacts -- periods that will likely span several million years, and be separated by periods of several tens of millions of years.

We have combined marine genera data, an orbital model of the Sun's path around the Milky Way with two face-on Galactic models. The first Galactic model is based on an artistic rendition of the Milky Way, by Churchwell et al. (2009). The second is an alteration of the first model, which accommodates all the extinctions within the spiral arms and displays a more symmetrical structure. Extinction data were then added to the new model and the existing orbital path of the Sun. In doing so all extinctions fall within the spiral arms.

Our new Galactic model, if correct, would support the idea that spiral arm crossings cause mass extinctions. Although a cyclic occurrence of large scale, global extinction is very likely to have an extraterrestrial origin, complex interactions at the Earth's surface cannot be completely discounted. It becomes harder to argue extinctions are caused by celestial events when they are not cyclic, which is still a controversial topic. 

However, reconsidering the possibility that spiral arms do cause extinctions on Earth, we can consider the ramifications of this to possible complex life elsewhere in the Galaxy. If we also consider that extinctions accelerate rather than impede the evolution of complex life, we may better constrain ideas on the Galactic Habitable Zone (Lineweaver et al. 2004). It may be the case that habitable planets orbiting stars further from the Galactic centre do not retain a high enough organic turn over rate due to more infrequent spiral arm crossings, or that closer in the extinction rate exceeds the ability for life to recover. Consideration could also be given to the possibility that habitable planet-hosting stars may have highly eccentric orbits around the Galaxy.    

Further statistical work and data on the structure of the Milky Way, kinematics, and the Solar orbit would refine our work and assist in continuing to test the spiral-arm/extinction hypothesis. Our future work will consist of two main threads -- the first being the consolidation of the archives of the Earth's impact history, extinction history, and the galactic architecture; the second being the construction of a detailed model that will allow us to test whether the Galactic structure is the dominant factor in defining the rate of Oort cloud comets (and hence impacts) at Earth.

This also lends itself to a prediction -- as our knowledge of the ancient Earth improves, if the hypothesis presented here is correct, then the periodicity should become clearer as more extinctions are found going further back. Is it reasonable to assume that the morphology of the Galaxy will have remained unchanged over the last four billion years. If so, then we could possibly argue that the periodic spiral arm crossings will have been happening all the way back -- albeit perhaps with some modulation on period and exact timings as a result of the evolution and disruption of the spiral arms and changes in the Sun's orbit around the galactic centre.


\acknowledgements{
We thank Scott Williams, Ivan Boji\v{c}i\'c and Graeme L. White for valuable discussion on this topic. We thank the referee for numerous helpful comments that have greatly improved the quality of this paper.
}



\references


Alroy, J.: 2008, \journal{Proc. Natl. Acad. Sci. U.S.A.}, \vol{105}, 11536.

Alvarez, L. W., Alvarez, W., Asaro, F. and Michel H. V.: 1980, \journal{Science}, \vol{208}, 1095.

Bailer-Jones, C. A. L.: 2009, \journal{Int. J. Astrobiology}, \vol{8}, 213.

Baldwin, E. C., Milner, D. J., Burchell, M. J. and Crawford, I. A.: 2007, \journal{Meteoritics Planet. Sci.}, \vol{42}, 1905.

Beech, M.: 2011, \journal{Astrophys. Space Sci.}, \vol{336}, 287.

Bottke, W. F., Vokrouhlick\'y, D. and Nesvorn\'y, D.: 2007, \journal{Nature}, \vol{449}, 48.

Chang, H.-Y. and Moon, H.-K.: 2005, \journal{Publ. Astron. Soc. Jpn.}, \vol{57}, 487.

Chapman, C. R., Cohen, B. A. and Grinspoon, D. H.: 2007, \journal{Icarus}, \vol{189}, 233.

Churchwell, E., Babler, B. L., Meade, M. R., Whitney, B. A., Benjamin, R., Indebetouw, R., Cyganowski, C., Robitaille, T. P., Povich, M., Watson, C. and Bracker, S.: 2009, \journal{Publ. Astron. Soc. Pac.}, \vol{121}, 213.

Davis, M., Hut, P. and Muller, R. A.: 1984, \journal{Nature}, \vol{308}, 715.

Feng, F. and Bailer-Jones, C. A. L.: 2013, \journal{Astrophys. J.}, \vol{768}, 152

Gies, D. R. and Helsel, J. W.: 2005, \journal{Astrophys. J.}, \vol{626}, 844. 

Grieve, R. A. F. and Pesonen, L. J.: 1992, \journal{Tectonophysics}, \vol{216}, 1.

Gogarten-Boekels, M., Hilario, E. and Gogarten, J. P.: 1995, \journal{Orig. Life Evol. Biosph.}, \vol{25}, 251.

Gomes, R., Levison, H. F., Tsiganis, K. and Morbidelli, A.: 2005, \journal{Nature}, \vol{435}, 466.

Hammel, H. B., Beebe, R. F., Ingersoll, A. P., Orton, G. S., Mills, J. R., Simon, A. A., Chodas, P., Clarke, J. T., de Jong, E., Dowling, T. E., Harrington, J., Huber, L. F., Karkoschka, E., Santori, C. M., Toigo, A., Yeomans, D. and West, R. A.: 1995, \journal{Science}, \vol{267}, 1288.

Haskin, L. A., Korotev, R. L., Rocklow, K. M. and Jolliff, B. L.: 1998, \journal{Meteoritics Planet. Sci.}, \vol{33}, 959.

Horner, J. and Evans, N. W.: 2002, \journal{Mon. Not. R. Astron. Soc.}, \vol{335}, 641.

Horner, J. and Jones, B. W.; 2010, \journal{Special issue Papers from the Astrobiology Society of Britain}, \vol{9}, 273.

Horner, J., Mousis, O., Petit, J.-M. and Jones, B. W.: 2009, \journal{Planet. Space Sci.}, \vol{57}, 338.

Horner, J. and Jones, B. W.; 2011, \journal{Astron. Geophys.}, \vol{52}, 1.16.

Levison, H. F., Morbidelli, A., Vanlaerhoven, C., Gomes, R. and Tsiganis, K.: 2008, \journal{Icarus}, \vol{196}, 258.

Lineweaver, C. H., Fenner, Y. and Gibson, B. K.: 2004, \journal{Science}, \vol{303}, 59.

Lykawka, P. S., Horner, J., Jones, B. W. and Mukai, T.: 2009, \journal{Mon. Not. R. Astron. Soc.}, \vol{398}, 1715.

Malhotra, R.: 1995, \journal{Astron. J.}, \vol{110}, 420. 

Matese, J. J., Whitman, P. G., Innanen, K. A. and Valtonen, M. J.: 1995, \journal{Icarus}, \vol{116}, 255.

Matese, J. J., Whitman, P. G. and Whitmire, D. P.: 1999, \journal{Icarus}, \vol{141}, 354.

Matese, J. J. and Whitmire, D. P.: 2011, \journal{Icarus}, \vol{211}, 926.

Mayhew, P. J., Jenkins, G. B. and Benton, T. G.: 2008, \journal{Proc. of the Roy. Soc. B.}, \vol{275}, 1630.

McElwain, J. C. and Punyasena, S. W.: 2007, \journal{Trends in Ecology and Evolution}, \vol{22}, 548.

McClure-GrifÞths, N. M., Dickey, J. M., Gaensler, B. M. and A. J. Green: 2004, \journal{Astrophys. J.}, \vol{607}, L127.

Melott, A. L. and Bambach, R. L.: 2011, Paleobiology, \vol{37}, 92.

Melott, A. L. and Bambach, R. L.: 2013, \journal{Astrophys. J.}, \vol{773}, 6.

Milner, D. J., Baldwin, E. C. and Burchell, M. J.: 2008, \journal{Meteoritics Planet. Sci.}, \vol{43}, 2015.

Minton, D. A. and Malhotra, R.: 2011, \journal{Astrophys. J.}, \vol{732}, 53.

Nesvorn\'y, D., Vokrouhlick\'y, D. and Morbidelli, A.: 2013, \journal{Astrophys. J.}, \vol{768}, 45.

Norman, M.: 2009, in: The lunar cataclysm: Reality or ''Mythconception''?, Elements, \vol{5}, 23.

Oberbeck, V. R. and Fogleman, G.: 1989, \journal{Nature}, \vol{339}, 434.

Oort, J. H.: 1950, \journal{Bull. Astron. Inst. Neth.}, \vol{11}, 91.

Overholt, A. C., Melott, A. L. and Pohl, M.: 2009, \journal {Astrophys. J.}, \vol{705}, L101.

Porto de Mello, G. F., L\'epine, J. R. and da Silva Dias, W.: 2009, \journal{Publ. Astron. Soc. Pac.}, \vol{420}, 349.

Rampino, M. R. and Stothers, R. B.: 1984, \journal{Nature}, \vol{308}, 709.

Rampino, M. R.: 1997, \journal{Celest. Mech. Dyn. Astron.}, \vol{69}, 49.

Raup, D. M. and Sepkoski, J. J.: 1986, \journal{Science}, \vol{231}, 833.

Rohde, R. A. and Muller, R. A.: 2005, \journal{Nature}, \vol{434}, 208.

S\'anchez-Lavega, A., Wesley, A., Orton, G., Hueso, R., Perez-Hoyos, S., Fletcher, L. N., Yanamandra-Fisher, P., Legarreta, J., de Pater, I., Hammel, H., Simon-Miller, A., Gomez-Forrellad, J. M., Ortiz, J. L., Garc'a-Melendo, E., Puetter, R. C. and Chodas, P.: 2010, \journal{Astrophys. J.}, \vol{715}, L155.

Sepkoski, J. J.: 2002, \journal{Bulletins of American Paleontology}, \vol{363}, 560.

Svensmark, H.: 2006, \journal{Astron. Nachr.}, \vol{327}, 866.

Svensmark, H.: 2012, \journal{Mon. Not. R. Astron. Soc.}, \vol{423}, 1234.

Sumi, T., et al.: 2011, \journal{Nature}, \vol{473}, 349.

Wells, L. E., Armstrong, J. C. and Gonzalez, G.: 2003, \journal{Icarus}, \vol{162}, 38.

Whitmire, D. P. and Jackson, A. A.: 1984, \journal{Nature}, \vol{308}, 713.

Wignall, P. D.: 2001, \journal{Earth-Science Reviews}, \vol{53}, 1.

\endreferences 
\end{multicols}

\vfill\eject

{\ }



\naslov{MASOVNA EKSTINKCIJA I STRUKTURA MLEQNOG PUTA}


\authors{M. D.~Filipovi\'c$^1$, J.~Horner$^{2,3}$, E. J. Crawford$^1$, N. F. H. Tothill$^1$ }

\vskip 3mm


\address{$^1$University of Western Sydney, Locked Bag 1797, Penrith South DC, NSW 1797, Australia}
\Email{m.filipovic}{uws.edu.au n.tothill@uws.edu.au e.crawford@uws.edu.au}
\address{$^2$School of Physics, University of New South Wales, Sydney 2052, Australia} 
\address{$^3$Australian Centre for Astrobiology, University of New South Wales, Sydney 2052, Australia}
\Email{j.a.horner}{unsw.edu.au}

\vskip.7cm


\centerline{UDK \udc}



\vskip.7cm

\begin{multicols}{2}
{


{\rrm Koriste{\cc}i najnoviji model Mleqnog Puta i orbite Sunca testirali smo hipotezu da svaki prolazak Sunca kroz spiralnu granu prouzrokuje masivne ekstinkcije svih {\zz}ivih organizama na Zemlji. Ovde predstav{\lj}amo novi model Mleqnog Puta koji je baziran na ma{nj}im modifikacijama da{lj}eg i te{\zz}e vid{lj}ivog dela na{\ss}e galaksije. Takodje, na{\ss} novi model smo prilagodili do sada poznatih i istorijski potvrdjenih {\ss}est (6) masovnih ekstinkcija. Uz ovih poznatih {\ss}est, predlo{\zz}ili smo jo{\ss} pet novih koji su proraqunati iz znaqajne redukcije podvodnih {\zz}ivih organizama u periodima od pre 415, 322, 300, 145 i 33 miliona godina.  Na{\ss}e simulacije pokazuju da je Sunce provelo najmanje 60\% vremena u spiralnim granama {\ss}to dodatno potvr{dj}uje povezanost masivnih ekstinkcija sa prolazom Sunca kroz spiralne grane Mleqnog Puta. Tako{dj}e, razmatramo i uticaj ovde predstav{lj}enog novog modela na{\ss}e Galaksije na stabilnost Galaktiqke Habitabilne Zone.
}

}
\end{multicols}

\end{document}